\documentclass[english,twocolumn,showpacs,prl,preprintnumbers,amsmath,amssymb]{revtex4}
\usepackage{graphicx}
\usepackage{epstopdf}
\usepackage{color}

\makeatletter


\newcommand{\aver}[1]{\langle {#1} \rangle}

\newcommand{\bc}[0]{b_{\mathrm{c}}}
\newcommand{\sce}[0]{\sigma_{\mathrm{ce}}}
\newcommand{\sL}[0]{\sigma_{\mathrm{L}}}
\newcommand{\Kce}[0]{K}
\newcommand{\Es}[0]{E_{\mathrm{s}}}

\def\c{s}
\def\d{s}
\definecolor{BrickRed}{cmyk}{0,0.89,0.94,0.28}
\definecolor{MidnightBlue}{cmyk}{0.98,0.13,0,0.43}
\definecolor{DarkGreen}{rgb}{0,0.7,0.1}

\newcommand{\comm}[1]{\if\c\d{{\color{MidnightBlue}\{\small \sc #1\}}}\else{}\fi}
\newcommand{\add}[1]{\if\c\d{{\color{magenta}#1}}\else{#1}\fi}
\newcommand{\drop}[1]{\if\c\d{{\color{DarkGreen}[[#1]]}}\else{}\fi}

\makeatother

\usepackage{babel}
\makeatother
\begin{document}

\preprint{XXX}

\title{Observation of Cold Collisions between Trapped Ions and Trapped Atoms}

\author{Andrew T. Grier}

\affiliation{Department of Physics, MIT-Harvard Center for Ultracold Atoms, and
Research Laboratory of Electronics, Massachusetts Institute of Technology,
Cambridge, Massachusetts 02139, USA}

\author{Marko Cetina}

\affiliation{Department of Physics, MIT-Harvard Center for Ultracold Atoms, and Research Laboratory of Electronics, Massachusetts Institute of Technology, Cambridge, Massachusetts 02139, USA}

\author{Fedja Oru\v{c}evi\'{c}}

\affiliation{Department of Physics, MIT-Harvard Center for Ultracold Atoms, and
Research Laboratory of Electronics, Massachusetts Institute of Technology,
Cambridge, Massachusetts 02139, USA}

\author{Vladan Vuleti\'{c}}

\affiliation{Department of Physics, MIT-Harvard Center for Ultracold Atoms, and
Research Laboratory of Electronics, Massachusetts Institute of Technology,
Cambridge, Massachusetts 02139, USA}

\date{\today}

\begin{abstract}

We study cold collisions between trapped ions and trapped atoms in the semiclassical (Langevin) regime. Using Yb$^+$ ions confined in a Paul trap and Yb atoms in a magneto-optical trap, we investigate charge-exchange collisions of several isotopes over three decades of collision energies down to 3~$\mu$eV ($k_B \times 35$~mK). The minimum measured rate coefficient of $6 \times 10^{-10}$~cm$^{3}\,\mathrm{s}^{-1}$ is in good agreement with that derived from a Langevin model for an atomic polarizability of $143$~a.u.

\end{abstract}


\maketitle

Studies of cold collisions between trapped neutral atoms have revealed a plethora of fascinating quantum phenomena, including Wigner threshold laws \cite{Kohler06}, magnetically tunable Feshbach resonances \cite{Inouye98}, controlled molecule formation \cite{Donley02}, and the suppression of individual scattering channels \cite{DeMarco99}. Collisions between trapped ions, on the other hand, are featureless, since the strong long-range repulsive Coulomb interaction prevents the ions from approaching each other. Collisions between ions and neutral atoms \cite{Langevin05,Cote00,Makarov03,Idziaszek09} fall into an intermediate regime where an attractive long-range $r^{-4}$ potential leads to semiclassical behavior for a wide range of collision energies, but where quantum phenomena dominate at very low energies. Cold ion-atom collisions have been proposed as a means to implement quantum gates \cite{Calarco07}, to cool atoms \cite{Makarov03,Moriwaki92} or molecules \cite{Smith05,Molhave00} lacking closed optical transitions, to bind small Bose-Einstein condensates to an ion \cite{Cote02}, or to demonstrate novel charge-transport dynamics \cite{Cote00a}.

As a function of collision energy, ion-atom collisions exhibit three distinct regimes \cite{Cote00,Makarov03}: a high-energy classical (hot) regime with a logarithmic dependence of cross section $\sigma$ on energy $E$, a wide semiclassical Langevin (cold) regime with a power-law dependence $\sigma(E) \propto E^{-1/2}$ for charge-exchange and momentum transfer processes \cite{Langevin05,Cote00}, and a quantum (ultracold) regime where contributions from individual partial waves can be distinguished. However, given the large forces on ions produced by small stray electric fields, it has been difficult to reach experimentally even the semiclassical regime.  Refs. \cite{Schuessler83,Hadjar04}, studying charge-exchange at $E \sim 100$~meV, report the only observations of Langevin-type ion-atom collisions. In ion-molecule systems, experimental signatures of Langevin collisions have been seen at high temperature \cite{Speck01,Kwong00,Motohashi05}, and recently also at 1~K (80~$\mu$eV) \cite{Willitsch08}. In all previous work, at most one of the collision partners was trapped.

In this Letter, we study collisions between independently trapped, laser-cooled ions and atoms down to unprecedented low energy (3~$\mu$eV) in the semiclassical collision regime. Using a double-trap system \cite{Cetina07}, we investigate resonant charge-exchange collisions for different Yb$^{+}+$Yb isotope combinations and find agreement with the Langevin model to within a factor of two over three decades of energy \cite{Langevin05,Cote00}. The highest energy 4~meV~=~$k_B \times 45$~K corresponds to the transition to the classical regime \cite{Zhang07a,ZhangPC}, while at the lowest energy 3~$\mu$eV~=~$k_B \times 35$~mK, where approximately 40 partial waves contribute to the cross section, isotope shifts should become relevant. The lower limit on collision energy is set by our ability to detect and minimize the micromotion of a single ion in the Paul trap.

The long-range interaction potential between a singly charged ion and a neutral atom is the energy of the induced atomic dipole in the ion's electric field, given by $V(r)=-C_4/(2r^4)$, where $C_4=\alpha q^2/(4 \pi \epsilon_0)^2$ is proportional to the atomic polarizability $\alpha$, and $q$ is the electron charge.  For given collision energy $E$ in the center-of-mass frame, there exists a critical impact parameter $\bc~=~(2 C_4 /E)^{1/4}$ that separates two types of collisions: those with impact parameter $b<\bc$ that result in inward-spiraling orbits of radius approaching zero, and those with $b>\bc$ that never cross the angular-momentum barrier \cite{Langevin05,Gioumousis58}.

For a resonant charge-exchange process $A^+ + A \to A+A^+$ of an ion with its parent atom, a semiclassical cross section $\sce$ can be simply derived, provided that $\bc$ is large compared to the range of the molecular potential: for collisions with $b>\bc$ the electron should remain bound to the incoming atom, while for inward-spiraling collisions with $b < \bc$ the electron should, with equal probability, attach to either nucleus. It follows that the resonant charge-exchange cross section is $\sce = \sL/2$, where $\sL= \pi \bc^2$ is the Langevin cross section. The corresponding semiclassical rate coefficient $K_{\mathrm{ce}} = \sce v =\pi \sqrt{C_4/ \mu}$, where $v$ is the relative velocity and $\mu$ the reduced mass, is independent of energy.

At high energies the model becomes invalid when $\bc$ becomes so small that charge exchange outside the centrifugal barrier starts to play a role; for Yb, this occurs at energies $E \gtrsim 10$~meV$ = k_B \times 120$~K \cite{Cote00,Zhang07a,ZhangPC}. At very low energies the semiclassical Langevin model ceases to be valid when the $s$-wave scattering limit is reached, which happens near $\Es \sim \hbar^4/(2 \mu^2 C_4) = 4$~peV~$= k_B \times 50$~nK. Quantum simulations \cite{Cote00} of this system \cite{Zhang07a} confirm the semiclassical Langevin model in the indicated energy range. For collisions between different isotopes, the Langevin expression should be modified at low collision energies when they become comparable to the small difference in binding energy of the electron to the two nuclei, i.e. the isotope shift of the ionization potential.  In this case we expect (exo-) endo-energetic charge exchange collisions to be (enhanced) suppressed \cite{Bodo08,Esry00}.

\begin{figure}
\includegraphics[width=3in]{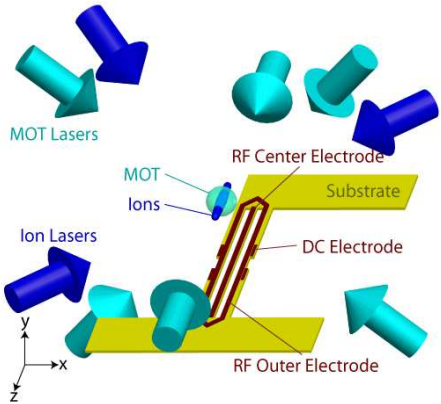}
\caption{Setup for trapping neutral (MOT) and singly charged (surface planar Paul trap) Yb in the same spatial volume.  Stray \emph{dc} field compensation and trapping along $\hat{z}$ for the Paul trap is provided by \emph{dc} electrodes. Imaging CCDs point along $(1,0,1)$ and $(0,1,1)$ axes. (color online) \label{setup} }
\end{figure}

We use a magneto-optical trap (MOT) in combination with a Paul trap, as originally proposed by W. Smith \cite{Smith05}. The setup, shown in Fig.~\ref{setup}, is described in detail elsewhere \cite{Cetina07}. In the present work, $^{172}$Yb, $^{174}$Yb, or $^{171}$Yb atoms are selectively loaded from an atomic beam into a MOT by tuning the laser frequency near the $^{1}$S$_{0}\to^{1}$P$_{1}$ transition at  $\lambda_a = 398.8$~nm. Typically, the MOT is operated at a 75 G/cm magnetic-field gradient and contains 3$\times$10$^{5}$ atoms at a peak density of $2\times10^{8}$~cm$^{-3}$ and temperature of 700~$\mu$K as determined by a time-of-flight measurement.

Cold ions are produced by nonresonant photoionization from the excited $^{1}$P$_{1}$ state of the MOT with 370-nm light from a secmiconductor laser \cite{Cetina07}. The ion trap is a surface-electrode Paul trap printed on a vacuum-compatible substrate \cite{Cetina07}, and is typically operated at 1.4~MHz to create a 0.3~eV deep pseudopotential trap 3.6~mm above the trap surface with a secular frequency of 67~kHz. Ion trap populations can be adjusted between a single and $10^{4}$ ions by varying the trap loading time.

Two beams from the same 370~nm laser used for photoionization provide Doppler cooling of the ions on the $^{2}$S$_{1/2}\to^{2}$P$_{1/2}$ transition along all three principal trap axes.  Ions that decay to a metastable D state are repumped \cite{Bell91} using a laser operating at 935~nm \cite{Cetina07}. We detect the ion population by monitoring trap fluorescence at 370~nm with a photomultiplier tube. A single cold trapped ion produces 5~kcounts/s.

\begin{figure}
\includegraphics[width=3.25in]{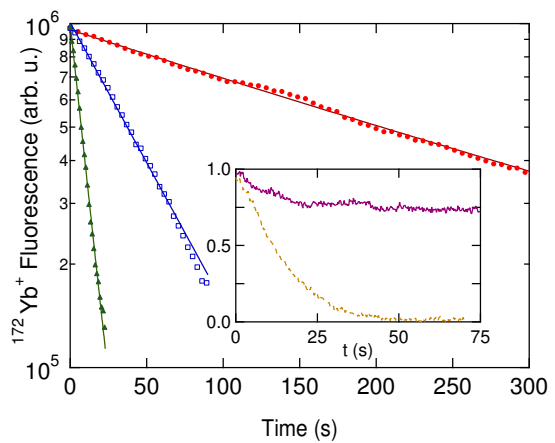}
\caption{Typical $^{172}$Yb$^{+}$ ion-crystal fluorescence decay and exponential fits for no (red circles), moderate (blue squares), and high (green triangles) $^{174}$Yb atomic density at the ion trap site.  Inset: $^{172}$Yb$^{+}$ (dashed, gold) and $^{174}$Yb$^{+}$ (solid, purple) population evolution in the presence of $^{174}$Yb.  Both traces are normalized by the same peak value. (color online)\label{Decay} }
\end{figure}

Collisions that only change the particles' energy and momentum are difficult to observe in our setup due to the continuous laser cooling. On the other hand, charge-exchange collisions between different isotopes $^\alpha$Yb$^+$ + $^\beta$Yb $\to ^\alpha$Yb + $^\beta$Yb$^+$ are easy to observe using isotope-selective ion fluorescence: we first load the ion trap from the MOT with isotope $^\alpha$Yb$^+$, change the MOT isotope to $^\beta$Yb by adjusting the frequency of the 399-nm laser, and then monitor the decay of the $^\alpha$Yb$^+$ ion population through the decay of the 370-nm fluorescence. In order to prevent photoionization and direct loading of the ion trap with $^\beta$Yb$^+$, we modulate the MOT and ion light out of phase to ensure that the MOT contains no excited atoms when the 370-nm light is present. Without the MOT, the ion trap loss is exponential with a typical lifetime $\tau_0 = 400$~s (Fig.~\ref{Decay}, red circles) for ion crystals, presumably due to collisions with background gas atoms. In the presence of the MOT, $\tau$ is substantially shortened, with a shorter lifetime for higher MOT density. The simple exponential decay over a decade in ion number indicates that the loss involves a single ion, rather than collisions between ions.  We also measure the trap fluorescence for identical ion and MOT isotopes. In the latter case, we observe a small initial decay, presumably due to a small amount of heating of the outermost ions, but no exponential loss (Fig.~\ref{Decay} inset). This confirms that the measured trap loss for different ion and MOT isotopes is indeed due to charge-exchange collisions, and not to heating in ion-atom collisions that interrupt the ion's micromotion \cite{Major68}.

\begin{figure}
\includegraphics[width=3.25in]{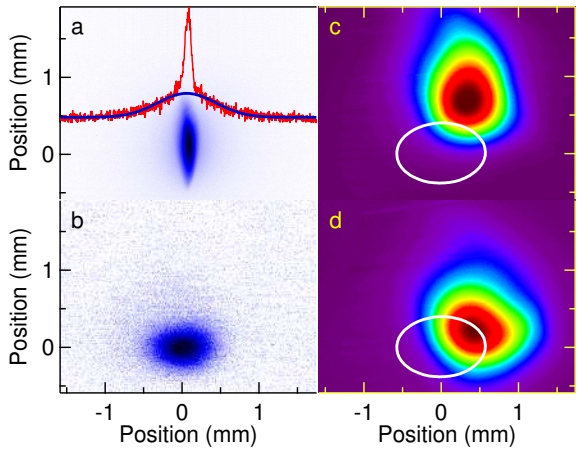}
\caption{(a) $(1,0,1)$ camera image of the ion crystal (blue) and cross-section showing highly non-Gaussian shape of crystal (red). (b) $(0,1,1)$ camera image of the ion crystal. (c) Typical low-overlap setting between MOT (colored contours) and $1/e^{2}$ contour of ions (white). (d) Same as (c) but for a higher overlap setting. (color online) \label{Images} }
\end{figure}

To accurately determine the charge-exchange rate coefficient, we vary the atomic density at the ion trap location by moving the zero of the MOT magnetic quadrupole field with a bias field. We determine the local atomic density at the ions' location by taking images of both the MOT cloud and the ions on charge-coupled device (CCD) cameras along two different directions (see Fig.~\ref{Images}), with the total atom number calibrated by means of a resonant-absorption measurement.  We then calculate the average atomic density experienced by the ions $\aver{n} = \int p(\mathbf{r}) n(\mathbf{r}) \textrm{d}\mathbf{r}$, where $p(\mathbf{r})$ is normalized ion distribution and $n(\mathbf{r})$ is the local atomic density. The inelastic rate coefficient $\Kce$ is obtained as the slope of a linear fit of the observed ion decay rate constant $\Gamma =1/\tau$ vs. $\aver{n}$.

We note that in most situations the ions' kinetic energy is determined by their micromotion throughout the \emph{rf} cycle \cite{Berkeland98}.  Any ion displacement from the zero of the oscillating electric field, be it due to ion crystal size or a $dc$ electric field, results in micromotion whose energy easily exceeds that of the thermal motion in the secular potential of the Paul trap. To investigate the dependence of the rate coefficient $\Kce$ on collision energy $E$, we vary the latter by loading different numbers of ions into the trap or by intentionally offsetting the ion crystal from the zero of the \emph{rf} quadrupole field with a \emph{dc} electric field.

\begin{figure}
\includegraphics[width=3.5in]{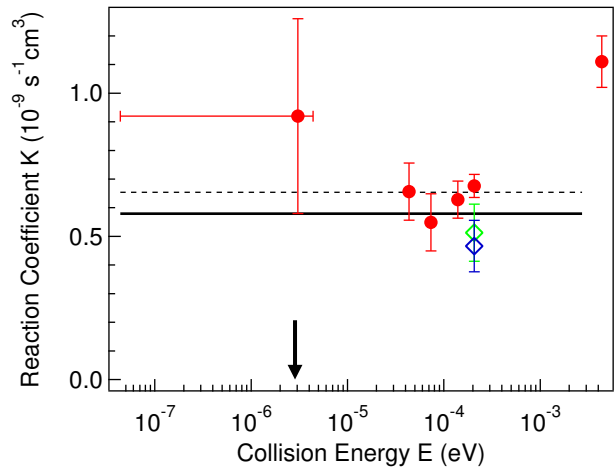}
\caption{Charge-exchange rate coefficient $\Kce$ as a function of collision energy in the center-of-mass frame. Red circles, green, and blue diamonds represent $^{172}$Yb$^{+}+^{174}$Yb, $^{172}$Yb$^{+}+^{171}$Yb, and $^{174}$Yb$^{+}+^{172}$Yb, respectively.  The solid line indicates the theoretical Langevin rate coefficient $K_{\textrm{ce}}$ \cite{Zhang07a},  the dashed line assumes a contribution from the the P-state polarizability \cite{Porsev99} (see text).  The black arrow indicates the ionization isotope shift between $^{174}$Yb and $^{172}$Yb.  (color online) \label{RateVEnergy} }
\end{figure}

The rate coefficient $\Kce(E)$ is shown in Fig.~\ref{RateVEnergy}. In the region $E \gtrsim 30$~$\mu$eV the ions' energy is calculated from the observed Doppler broadening. For well-compensated traps containing small crystals or single ions, where the Doppler broadening is smaller than the natural linewidth, we determine the average micromotion energy from the observed correlation between \emph{rf} drive signal and ion fluorescence \cite{Berkeland98}. In the latter case, the signal-to-noise ratio is small, resulting in larger energy uncertainty. For all energies investigated here, the atoms' contribution to the collision energy is negligible.

The data point at the lowest energy $E=3.1$~$\mu$eV, being an average of 22 single-ion measurements, has large uncertainty in $K$. It also has large energy uncertainty because the micromotion is sufficiently well compensated for the fluorescence correlation signal to be dominated by noise. The other data points show observable Doppler broadening and were measured with up to a few thousand ions, resulting in smaller statistical uncertainties. We estimate systematic uncertainties to be a factor of two in $\Kce$ due to the difficulty of absolute MOT density calibration, and $\pm 50\%$ in $E$, due to the non-thermal energy distribution of the micromotion.

The \emph{ab initio} calculated value for the polarizability $\alpha_\textrm{S}$~=143~a.u. = $ \epsilon_{0} \cdot 2.66\times10^{-28}$~m$^{3}$ \cite{Zhang07a} for the atomic $^1$S$_0$ ground state yields $K_{\textrm{ce}} = 5.8 \times 10^{-10}$~cm$^{3}\,\mathrm{s}^{-1}$ for ground-state collisions (solid line in Fig.~\ref{RateVEnergy}). Our experimentally measured value of $K = 6 \times 10^{-10}$~cm$^{3}\,\mathrm{s}^{-1}$ in the middle of the semiclassical region \cite{Zhang07a} around $E\sim$100~$\mu$eV is thus in good agreement with the Langevin model. The data point at the highest collision energy $E = 4$~meV coincides with the transition to the classical region, where the rate coefficient increases with collision energy \cite{Zhang07a,ZhangPC}. Doubly excited collisions Yb$^{+*}$+Yb$^*$ cannot occur as the excitation light for ions and atoms is modulated out of phase. For Yb$^{+*}$+Yb collisions, the Langevin cross section, depending only on the ion's charge and the atom's polarizability, is unchanged, and the charge-exchange process remains resonant, yielding the same value of $\sce$. Yb$^{+}$+Yb$^*$ collisions cannot contribute for our collision energies since the atomic transition is tuned out of resonance with the MOT light at an ion-atom distance $R_0\approx 30$~nm larger than $\bc = 2-12$~nm, and modeling of collision trajectories shows that the time required to move between $r=R_0$ and $r=\bc$ exceeds several excited-state lifetimes. For illustration purposes, we have indicated in Fig.~\ref{RateVEnergy} the quite small change in $K_{\textrm{ce}}$ if Yb$^{+}$+Yb$^*$ collisions with $\alpha_{\textrm{P}}$~=~500~a.u. for the $^1$P$_1$ state \cite{Porsev99} were to contribute at our measured MOT excited-state fraction.

The theoretical value $K_{\textrm{ce}}=\sigma_{\textrm{L}} v/2$, corresponding to equal binding probability of the electron to the two nuclei, should apply when the collision energy far exceeds the isotope shift of the ionization potential. The latter can be estimated from spectroscopic data of a transition to a state with low electron probability density at the nucleus, such as $4f^{14}6s^{2} \to 4f^{14}6s10d$ \cite{Kischkel92}. It follows that the $^{172}$Yb$^{+}+^{174}$Yb $\to$ $^{172}$Yb$+^{174}$Yb$^{+}$ reaction should be exothermic and release $\Delta E$ = 2.9~$\mu$eV = $h \times 0.7$~GHz, and other isotope combinations give similar values. At the substantially larger collision energy $E$ = 0.21~meV $ \gg \Delta E$ we have investigated various isotope combinations, $^{172}$Yb$^{+} + ^{174}$Yb, $^{174}$Yb$^{+} + ^{172}$Yb, and $^{174}$Yb$^{+} + ^{171}$Yb, and we indeed find that they they all display the same rate coefficient (see Fig.~\ref{RateVEnergy}). For the data point $^{172}$Yb$^{+}+^{174}$Yb $\to$ $^{172}$Yb$+^{174}$Yb$^{+}$ at the lowest energy $E$ = 3.1~$\mu$eV $\approx$ $\Delta E$, we speculate, based on calculations for similar processes in H \cite{Bodo08,Esry00}, that the exothermic collision rate should exceed the resonant Langevin value, which is (weakly) supported by our data.

While it may be impossible to compensate stray fields well enough to reach the $s$-wave scattering limit $\Es = 4$~peV, Feshbach \cite{Idziaszek09} and other collision resonances may be observable well above $\Es$ \cite{Cote00,Zhang07a}. All the Yb isotopes used here have been cooled to quantum degeneracy in an optical dipole trap \cite{Takasu03,Fukuhara07}, which would allow the investigation of collision processes between an ion and a Bose-Einstein condensate or Fermi gas.  Alternatively, to avoid the large resonant charge-exchange cross section observed here, a different species such as Rb could be used for sympathetic cooling of ions \cite{Makarov03,Moriwaki92}, or for studying ion impurities in a Bose-Einstein condensate \cite{Cote02}.

We would like to thank T. Pruttivarasin for technical assistance and D. DeMille, D. Leibfried, A. Dalgarno, P. Zhang, R. C\^ot\'e, W. Smith, and I. Chuang for suggestions and stimulating discussions.  This work was supported by the NSF and the NSF Center for Ultracold Atoms.


\end{document}